\documentclass[usenatbib]{mn2e}    
\usepackage{mathptmx}
\usepackage{graphicx}
\usepackage{float}
\usepackage{fixltx2e}
\usepackage{natbib}
\usepackage{amsmath}
\usepackage{amssymb}
\usepackage[usenames,dvips]{color}
\usepackage{appendix}

\newcommand{\msun}{M_\odot}

\title[Onset of environmental influence]{Why does the environmental influence on group and cluster galaxies extend beyond the virial radius?}
\author[Y.M. Bah\'{e} et al.]{Yannick~M.~Bah\'{e}$^1$\thanks{ybahe@ast.cam.ac.uk}, Ian G.~McCarthy$^{2,1}$, Michael L.~Balogh$^3$ and Andreea S. Font$^{2,1}$\\
$^1$ Institute of Astronomy, University of Cambridge, Madingley Road, Cambridge CB3 0HA, United Kingdom\\
$^2$ School of Physics and Astronomy, University of Birmingham, Edgbaston, Birmingham B15 2TT, United Kingdom\\
$^3$ Department of Physics and Astronomy, University of Waterloo, Waterloo, Ontario N2L 3G1, Canada
}
\date{Accepted 2013 January 16. Received 2013 January 7; in original form 2012 October 31}

\begin{document}
\label{firstpage}
\maketitle

\begin{abstract}
In the local Universe, galaxies in groups and clusters contain less gas and are less likely to be forming stars than their field counterparts. This effect is not limited to the central group/cluster regions, but is shown by recent observations to persist out to several virial radii. To gain insight into the extent and cause of this large-scale environmental influence, we use a suite of high-resolution cosmological hydrodynamic simulations to analyse galaxies around simulated groups and clusters of a wide range of mass (log$_{10}$ M$_\text{host}$/M$_\odot$ = [13.0, 15.2]).  In qualitative agreement with the observations, we find a systematic depletion of both hot and cold gas and a decline in the star forming fraction of galaxies as far out as $\sim$ 5 r$_{200}$ from the host centre.  A substantial fraction of these galaxies are on highly elliptical orbits and are not infalling for the first time ($\sim$ 50 per cent at 2 r$_{200}$, independent of host mass) or are affected by `pre-processing' (less than 10 per cent of galaxies around groups, increasing to $\sim$ 50 per cent near a massive cluster), even a combination of these indirect mechanisms does not fully account for the environmental influence, particularly in the case of the hot gas content.  Direct ram pressure interaction with an extended gas `halo' surrounding groups and clusters is shown to be sufficiently strong to strip the hot gas atmospheres of infalling galaxies out to $\sim$ 5 r$_{200}$.  We show that this influence is highly anisotropic, with ram pressure along filaments enhanced by up to a factor of 100 despite significant co-flow of gas and galaxies. 
\end{abstract}


\section{Introduction}
\label{sec:introduction}

There is strong observational evidence that the internal properties of galaxies depend on their local environment. One well-known example is the morphology--density relation \citep{Dressler_1980} by which early-type galaxies are more common in high-density environments such as the central regions of clusters, whereas late-type galaxies dominate the field population.  While much of this apparent trend is explained by galaxies in groups and clusters being typically more massive than field galaxies, combined with a correlation between galaxy mass and morphology in the field, there is still a significant difference in the morphologies of galaxies of {\it fixed stellar mass} in the field and those in massive groups and clusters (e.g., \citealt{Kauffmann_et_al_2004,Blanton_et_al_2005}).
A similar relation holds for galaxy colours: those in denser regions are preferentially redder than their isolated counterparts (e.g., \citealt{Hogg_et_al_2004}). As the colour of a galaxy is primarily influenced by its recent star formation activity, this `colour--density relation' indicates a reduced level of star formation in group and cluster galaxies compared to those of similar stellar mass in the field (e.g., \citealt{Balogh_et_al_2004,Poggianti_et_al_2006}). 

In principle, both of these trends may either be due to intrinsic differences between field and group/cluster galaxies (i.e., galaxies form differently in close proximity to a massive cluster, or a large-scale overdensity destined to become a cluster) or they may be the result of a transformation of late-type, star-forming, blue field galaxies into early-type, passive, red ones after they are accreted by a group or cluster. There are several mechanims that could trigger such transformations. The tidal field of the group or cluster, as well as interactions with other galaxies, which are naturally more common in dense environments may strip, re-shape, or even totally disrupt a galaxy (e.g., \citealt{Moore_et_al_1999}). At the same time, the high velocity of a galaxy relative to the intra-group/-cluster medium (ICM)\footnote{We use the term `ICM' for hot gas in and around groups/clusters which is not bound to infalling galaxies or sub-groups. This includes gas beyond the virial radius.} gives rise to ram pressure which can remove its cold gas reservoir (so-called ram pressure stripping, \citealt{Gunn_Gott_1972,Abadi_et_al_1999}) and therefore shut down star formation. The \emph{hot} gas haloes surrounding infalling galaxies are even easier to remove since these are much more extended, less dense and therefore less tightly bound to the galaxy (e.g., \citealt{Larson_et_al_1980,Balogh_et_al_2000,McCarthy_et_al_2008}). While not impacting star formation directly, the removal of hot gas does put an end to the replenishment (through cooling) of the cold gas used up in it. The result is a delayed decrease in star formation as the remaining cold gas is consumed (`strangulation' or `starvation'). The common theme of these mechanisms is that the environmental influence on galaxies falling into a group or cluster decreases with increasing distance from the centre, as the density of both the ICM and galaxies, as well as their orbital velocities, decrease.  It is less clear, however, at which point during a galaxy's infall one should expect these mechanisms to first `switch on'.  This depends on the detailed structural properties of the infalling galaxies as well as that of the host groups and clusters into which they are falling.

There is ample evidence from both observations and simulations that groups and clusters have no sharp `edge' to mark the transition to the field environment.  Instead, their outer regions blend smoothly into the surrounding large-scale structure. There is therefore no obvious starting point for the above-mentioned mechanisms to begin acting on galaxies.  A commonly employed boundary radius for a halo is the `virial radius' (hereafter r$_{200}$), which is often computed as the radius inside which the average density equals 200 times the critical density.  Roughly speaking, this radius corresponds to the extent of the virialised region of a halo in cosmological simulations.  There is, however, mounting observational evidence that the colour--density relation persists out to distances significantly greater than r$_{200}$ (e.g., \citealt{Balogh_et_al_1999,Hansen_et_al_2009,Haines_et_al_2009,von_der_Linden_et_al_2010,Lu_et_al_2012,Wetzel_et_al_2012,Rasmussen_et_al_2012}; see also \citealt{Bahe_et_al_2012}). If environmentally-induced transformations are indeed responsible for the observed trends, then galaxies are evidently affected well beyond the virial radius. 

One commonly identified \emph{indirect} way in which galaxies can be environmentally affected at large distances from the centre of a galaxy cluster is through `pre-processing' in infalling groups (e.g., \citealt{Berrier_et_al_2009,McGee_et_al_2009}). A second indirect mechanism is what we refer to as `overshooting': a non-negligible fraction of infalling galaxies are on highly elliptical orbits (e.g., \citealt{Benson_2005}) that bring them well within r$_{200}$ on first pericentric passage but back out beyond this radius later on (e.g., \citealt{Gill_et_al_2005,Ludlow_et_al_2009}).  The existence of environmental trends out to radii well beyond r$_{200}$ is therefore not necessarily incompatible with direct environmental influence being confined to smaller scales. 

Of course, the existence of these indirect mechanisms does not rule out the possibility that there is also direct environmental influence of the group or cluster at distances beyond the virial radius. As groups and clusters blend into the large-scale surrounding environment, a galaxy is surrounded by gas even at large distances from the host centre (see, e.g., \citealt{Frenk_et_al_1999} who show that the hot gas haloes of simulated massive galaxy clusters extend out as far as $\sim10$ Mpc) This gas will exert a ram pressure force on any galaxy moving relative to it, which may be sufficient to remove (some of) its gas. 

It is therefore conceivable that all three of these mechanisms are operating in concert to reduce the star formation activity of galaxies.  However, separating them from each other would be a formidable challenge for current observations, as there is in general only limited information about a galaxy's velocity and that of its surrounding gas.  Furthermore, it is extremely difficult to even {\it detect} the hot gas haloes of the nearest galaxies \citep{Bregman_2007}, let alone measure their structural properties around more distant galaxies falling into groups and clusters. 

Numerical simulations are a potentially promising tool to gain further insight, as both velocities and galaxy orbital histories are readily available.  The demands on these simulations, however, are considerable. Cosmological initial conditions are required for realistic galaxies and hosts, and to give meaningful information on ram pressure, baryons have to be included in the simulations directly, along with realistic physical prescriptions for relevant processes that affect the baryons, including radiative cooling, star formation, chemodynamics, and supernova feedback. The resolution must be high enough to resolve individual galaxies, while the simulation must also include rare objects such as massive galaxy clusters. High-resolution cosmological hydrodynamic simulations of large volumes would satisfy these requirements but their computational cost is currently prohibitively high. A promising compromise is the use of simulations with `zoomed' initial conditions, where a small region of a large, low-resolution simulation box is re-simulated at high resolution (e.g., \citealt{Tormen_et_al_1997}). 

In this paper, we use a set of simulations following this philosophy, the \textsc{Galaxies-Intergalactic Medium Interaction Caculations} (GIMIC; \citealt{Crain_et_al_2009}). By re-simulating selected regions of the Millennium Simulation (\citealt{Springel_et_al_2005}; which includes dark matter only), chosen to encompass a wide range of large-scale environments they include both sparse voids and a massive galaxy cluster, as well as numerous less massive clusters and galaxy groups. A particular advantage of \textsc{gimic} is that, over a relatively wide range of stellar mass, the simulated \emph{field} disc galaxies have already been shown to have properties in good agreement with a variety of observational data, including the relations between stellar mass and rotation velocity, size, and star formation efficiency (defined as the ratio of stellar mass to total mass; see \citealt{McCarthy_et_al_2012b}).  The simulations also reproduce the observed scalings of hot gas X-ray luminosity with K-band luminosity, star-formation rate and rotation velocity \citep{Crain_et_al_2010a}, as well as the properties of stellar haloes around Milky Way-mass disc galaxies (\citealt{Font_et_al_2011,McCarthy_et_al_2012a}).  With a reasonably realistic population of field galaxies, these simulations are therefore suitable to investigate the processes acting on them upon infall into a group or cluster. 

This paper is structured as follows. In Section 2, we briefly describe the simulations and our method for identifying and tracing galaxies within them. The extent of environmental influence on our simulated galaxies beyond r$_{200}$ is shown in Section 3, followed by an in-depth analysis of the underlying physical mechanisms in Section 4. In Section 5 we investigate the influence of filaments, before presenting our conclusions in Section 6. All masses and distances are given in physical units unless otherwise specified. A flat $\Lambda$CDM cosmology with Hubble parameter $h =$ H$_{0}/(100\,{\rm km}\,{\rm s}^{-1}{\rm Mpc}^{-1}) = 0.73$, dark energy density parameter $\Omega_\Lambda = 0.75$ (dark energy equation of state parameter $w=-1$), and matter density parameter $\Omega_{\rm M} = 0.25$ is used throughout this paper.

\section{Simulations and Analysis}
\label{sec:simulations}

\subsection{GIMIC simulations}
This work is based on the \textsc{Galaxies-Intergalactic Medium Interaction Calculation} suite of simulations (\textsc{gimic}). The reader is referred to \citet[see also \citealt{Schaye_et_al_2010}]{Crain_et_al_2009} for a full description of these simulations; here we only summarise their main features relevant to this study. 

The \textsc{gimic} simulations are a set of five re-simulations of nearly spherical regions of varying mean density extracted from the Millennium Simulation \citep{Springel_et_al_2005}. The regions are chosen so that at z = 1.5 their average densities differ from the cosmic mean by (-2, -1, 0, +1, +2) $\sigma$, where $\sigma$ is the rms mass fluctuation on a scale of $18 h^{-1}$ Mpc at this redshift. In this way, \textsc{gimic} includes rare objects such as a sparse void and, of particular importance here, many groups and clusters of galaxies, including a particularly massive one, with $\log_{10}$ (M$_{200}$ / M$_\odot$) $\approx 15.2$ at $z=0$ at the center of the +2$\sigma$ sphere.  

The simulations were carried out at 3 different resolutions: `low', `intermediate', and `high'. The `low' resolution is the same as in the original Millennium Simulation while the `intermediate' and `high' resolution simulations have $8$ and $64$ times better mass resolution, respectively.  As only the $-2\sigma$ and $0\sigma$ regions have been run at high resolution (owing to prohibitive computational expense), we use the intermediate-resolution simulations here.  These simulations have a baryon particle mass resolution of $m_\text{gas} \sim 1.16 \times 10^7 h^{-1} \msun$ with a gravitational softening that is 1 $h^{-1}$ kpc in physical space at $z \le 3$ and is fixed in comoving space at higher redshifts.  Thus, even relatively low-mass galaxies (M$_* \sim$ a few $10^9$ M$_\odot$) are resolved into several hundred particles, making \textsc{gimic} suitable to study the interaction between galaxies and groups/clusters of a wide range of masses.  We note that \citet{McCarthy_et_al_2012b} have shown that the star formation efficiencies and sizes of the simulated galaxies in \textsc{gimic} are approximately converged when there are (at least) several hundred star particles present (although a larger number is required before the $z=0$ specific star formation rates converge, see Section 3 for further discussion), while \citet{McCarthy_et_al_2008} have shown that the stripping of hot gas is converged when there are a similar number of hot gas particles present initially.

The simulations were carried out with the TreePM-SPH code \textsc{GADGET-3} (last described in \citealt{Springel_2005}) and include significantly modified prescriptions for star formation \citep{Schaye_DallaVecchia_2008}, metal-dependent radiative cooling in the presence of a \citet{Haardt_Madau_2001} UV/X-ray background \citep{Wiersma_et_al_2009a}, feedback and mass transport by Type Ia and Type II supernovae \citep{DallaVecchia_Schaye_2008}, as well as stellar evolution and chemodynamics \citep{Wiersma_et_al_2009b}.  However, they do not include a prescription for feedback due to accreting supermassive black holes, so that massive galaxies with $\log_{10} (\text{M}_* / \text{M}_\odot) \ga 10.7$ in \textsc{gimic} suffer from `over-cooling' (see \citealt{Crain_et_al_2009} and \citealt{McCarthy_et_al_2012b}).  As a result, we do not expect realistic predictions for massive galaxies and therefore limit our analysis to the range $\log_{10} (\text{M}_* / \text{M}_\odot) = [9.0, 11.0]$.  We note that the neglect of AGN feedback in these simulations also means that the {\it central} regions of the simulated groups and clusters will not be realistic in terms of the ICM properties or the size/mass of the central brightest galaxy \citep{McCarthy_et_al_2010}.  However, at very large distances from the group/cluster centre, the region on which we focus in this study, the cooling time of the ICM is much longer than a Hubble time, so that this does not affect the validity of our results.

\subsection{Selection and tracing of host groups and clusters}

Host groups and clusters of galaxies were identified at redshift $z$ = 0 using a standard Friends-of-Friends (FoF) algorithm with a linking length of $b$ = 0.2 times the mean inter-particle separation.  We select as hosts all FoF groups with M$_\text{bound} > 10^{13.0}$ M$_\odot$, where M$_\text{bound}$ is the mass of all gravitationally bound particles within this FoF group as identified by the \textsc{subfind} algorithm of \citet{Dolag_et_al_2009}. This version extends the standard implementation of \citet{Springel_et_al_2001} by including baryonic particles in the identification of bound substructures and also allows one to distinguish substructures which are located within still larger substructures (i.e., sub-subhaloes, sub-sub-subhaloes etc.) from those which are associated with the main subhalo of a FoF group.

This type of host mass threshold is somewhat different from the more commonly used spherical overdensity mass M$_{200}$. The reason for our choice is that we have identified many instances where the FoF algorithm will link together, e.g., a cluster-mass halo and a nearby infalling group-mass halo (see Section \ref{sec:preprocessing}).  In this case, no value of M$_{200}$ would be computed for the group-mass halo, since it is part of the overall FoF group, while M$_\text{bound}$ is still well-defined. As an additional benefit, this threshold excludes haloes which are affected by the presence of low-resolution particles near the edge of a simulation sphere and typically show extremely low values of M$_\text{bound}$.

\begin{figure}
\includegraphics[width=\columnwidth]{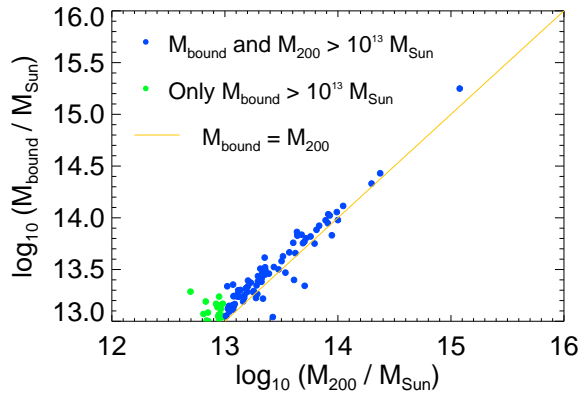}
\caption{Comparison between mass within r$_{200}$ (`M$_{200}$') and total gravitationally self-bound mass (`M$_{\text{bound}}$') of the host haloes used in this work. For most systems, there is very close agreement between these two, so that the host would be included in our sample with either definition (blue points). In general, M$_{\text{bound}}$ is slightly larger than M$_{200}$ due to bound structures extending beyond r$_{200}$, so a small number of systems is only included when using the former criterion (green points).}
\label{fig:hostmasses}
\end{figure}

In Fig.~\ref{fig:hostmasses} we present a comparison between M$_\text{bound}$ and M$_{200}$ for our host FoF haloes.  Most of them trace a narrow sequence with M$_\text{bound} \approx$ M$_{200}$ and so are selected irrespective of which type of mass cut is applied (blue points).  The former is slightly higher on average, due to bound structures extending beyond r$_{200}$, so there is a small set of galaxy-group scale haloes that are only included in our sample adopting the M$_\text{bound}$ cut (green points). In total, our sample includes $\sim 100$ host systems with masses in the range $13.0 \leq \log_{10}$ M$_\text{bound} / $ M$_\odot \la 15.2$.

To identify the \emph{progenitors} of the host haloes in previous snapshots, we use all the gravitationally bound dark matter particles in each FoF group in our $z = 0$ host sample as a tracer population. Using their unique particle IDs we identify the FoF halo to which the majority of the tracer particles belonged in the previous snapshot, which we designate as the progenitor.  We repeat this tracing procedure back to redshift $z = 10$.

\subsection{Galaxy identification and selection}

Having identified the host haloes and their progenitors in each simulation, we next select `galaxies' (i.e., self-bound subhaloes) in a similar way. Starting at the \emph{highest} redshift ($z = 10$), we identify for each galaxy detected by \textsc{subfind} its constituent dark matter and star particles. In the subsequent (`target') snapshot, we then search for the galaxy containing most of the mass of these particles and identify it as the original galaxy's descendent.  Any subhalo in the target snapshot which is not identified as a descendent is taken as the starting point for a new galaxy and the process is repeated until reaching the snapshot at $z = 0$. 

In the case that two or more galaxies have the same descendent in the target snapshot, we continue tracing only the one contributing the most mass; all others are marked as `accreted' onto this galaxy and are not traced further so that no galaxy is counted twice.

We also take into account the possibility that a galaxy may temporarily not be identified as gravitationally self-bound by \textsc{subfind} (e.g., \citealt{Muldrew_et_al_2011}). If a galaxy in snapshot $i$ has no descendent in snapshot $i+1$, or an identified descendent accounts for less than 50 per cent of its mass, we repeat our seach in snapshot $i+2$ and, provided a descendent is identified in this snapshot, we continue tracing the galaxy from there. This is particularly important for galaxies moving through the central regions of galaxy clusters where the high background density makes an erroneous non-detection more likely. 

To ensure that we can trace a galaxy falling into a group or cluster for as long as possible, we exclude all particles belonging to a host when identifying the descendent galaxies.  In this way, even if the vast majority of the galaxy's particles has been stripped, we will then still identify the subhalo made up from the remaining bound particles as its descendent, and not the host halo, which would only be the appropriate choice if the galaxy had been totally disrupted (in practical terms this means that there are less than 20 bound particles remaining, at which point \textsc{subfind} no longer classifies a structure as a self-bound subhalo). 

Finally, we select for analysis those galaxies whose total (bound) stellar mass in at least one snapshot falls within the range $\log_{10}$(M$_*$ / M$_\odot$) = [9.0, 11.0].  This results in a final sample of $\sim 30\, 000$ unique galaxies over all redshifts.  From these, we create two sub-sets: our main sample of galaxies in the vicinity of hosts (the `infall' sample) is formed by those that are found within 5 r$_{200}$ from the centre of a host in at least one snapshot; there are $\sim 15\, 000$ galaxies in this set.  For comparison purposes, we also form a sample of `field' galaxies, defined as centrals which never come closer than 5 r$_{200}$ to an FoF group with M$_{\text{bound}} \geq 10^{13}$ M$_\odot$. 

Our reason for identifying hosts at $z = 0$ and then tracing them backwards in time, while tracing galaxies forwards from $z = 10$ without any prior selection, is two-fold: on the one hand, being interested in the influence on galaxies by their host environment, we want to select only those hosts that are themselves evolving relatively undisturbed, without being accreted onto other, more massive hosts. Identification at $z = 0$ and then tracing backwards these `surviving hosts' satisfies this purpose. For galaxies, on the other hand, we also want to include those identified at $z > 0$ which have been disrupted or merged later on, and take into account the fact that the stellar mass of a galaxy may vary significantly over cosmic timescales. For this reason, we have chosen to trace them forward in time, and only select the actual `galaxy' subhaloes afterwards.  

\section{Large-scale environmental trends at z $\sim$ 0}
\label{sec:onset}

Within our traced sample of galaxies and hosts as described above, we now look at the extent of environmental influence on satellite galaxies.  For the purposes of the present study we focus on the fraction of galaxies which are star forming (defined as having a specific star formation rate SFR/M$_* \geq 10^{-11}$ yr$^{-1}$) and have hot or cold gas mass fractions M$_\text{gas}$/M$_*$ exceeding a threshold value of 0.1. We intend to explore a wider range of properties, including colours and various measures of morphology, in a future study. 

\begin{figure*}
\includegraphics[width=2\columnwidth]{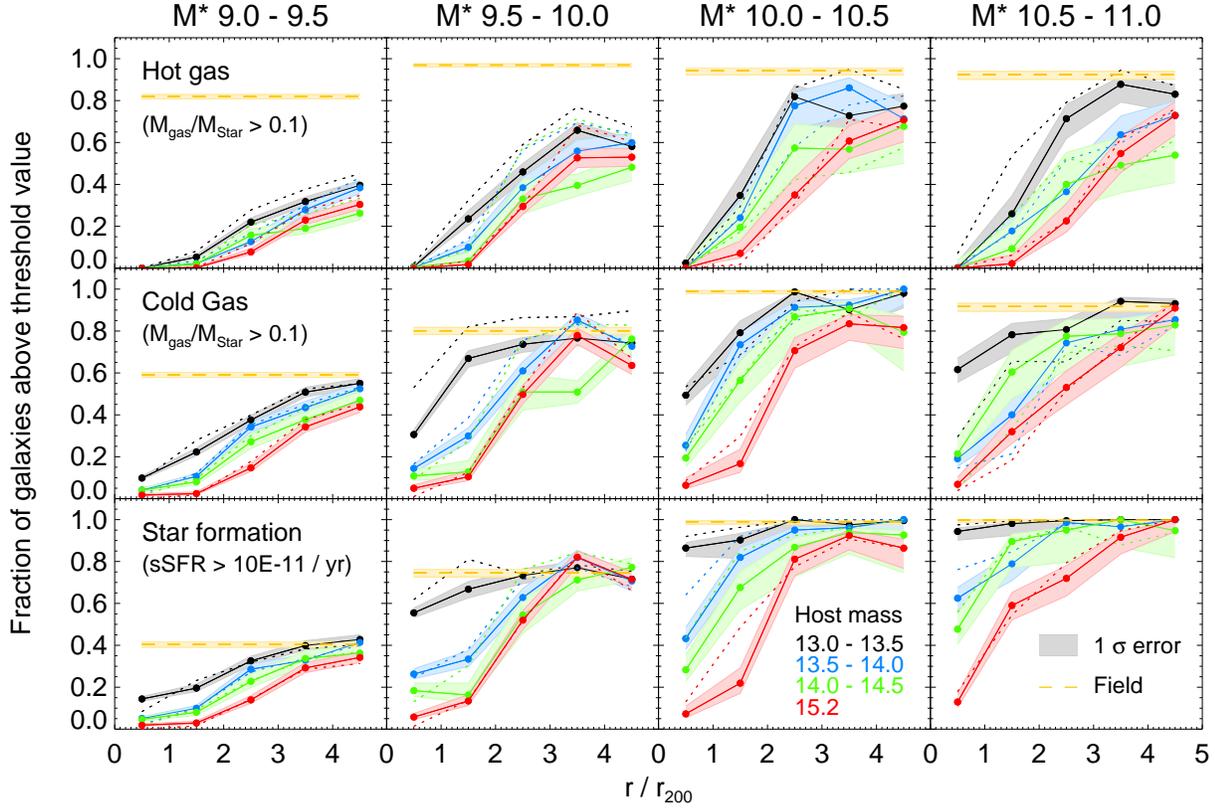}
\caption{Fraction of galaxies with large hot (top row) and cold gas content (middle row), and high specific star formation rates (bottom row) respectively at $z \sim 0$. Panels are split by stellar mass as given at the top; differently coloured lines represent hosts of different mass as indicated in the figure.  The shaded regions represent statistical $1\sigma$ binomial uncertainties, see text for details. Dotted lines show the results obtained when galaxies are binned by stellar mass at infall (when crossing 5 r$_{200}$ for the first time), as opposed to our standard method where we bin by stellar mass at the point when the galaxy is `observed' --- there is little difference between the two. The dashed yellow lines represent the corresponding fractions in the field sample. There are clear trends towards less gas and lower star forming likelihood well beyond r$_{200}$.}
\label{fig:onset}
\end{figure*}

In Fig.~\ref{fig:onset} we show the fraction of galaxies in which the ratio of total mass of gravitationally bound hot and cold gas to stellar mass (i.e., the hot and cold gas mass fraction, respectively) lies above a threshold value of $0.1$, as well as the fraction of galaxies with a specific star formation rate (sSFR) above $10^{-11}$ yr$^{-1}$.  Both of these thresholds are set by the resolution of our simulations and the desire to have a fixed (specific) threshold across the entire range of stellar masses that we explore.  Conveniently, the sSFR threshold that we adopt is very similar to that which is employed in many observational studies: these show a well-defined star-forming sequence, which is isolated from passive galaxies by an sSFR cut at 10$^{-11}$ yr$^{-1}$ (e.g., \citealt{Wetzel_et_al_2012}). To distinguish hot and cold gas, we adopt a threshold temperature of $T = 2 \times 10^5$ K; for cold gas we additionally require a density $n \geq 0.01$ cm$^{-3}$. Such a cut in density and temperature for the cold gas roughly mimics a selection of atomic (HI) gas. 

Statistical uncertainties, indicated by shaded bands in Fig.~\ref{fig:onset}, are calculated as binomial 1$\sigma$ confidence intervals using a beta distribution generator as proposed by \citet{Cameron_2011}. However, our stacking approach makes it possible for one galaxy to produce multiple data points in the same (radial) bin, which are clearly not independent of each other. To avoid underestimating our uncertainties because of this, we divide in each bin both the total number of data points, $n$, and the number of data points which lie above our chosen threshold, $k$, by the ratio $f = n/g$, where $g$ is the number of unique galaxies associated with the $n$ data points in this bin. We then calculate confidence intervals based on $k/f$ `successes' out of $g = n/f$ `trials' (instead of $k$ out of $n$), and thus ensure that our uncertainties are based on the number of \emph{independent} data points per bin. The same approach is adopted for all other plots in this paper which show the fraction of galaxies satisfying a particular criterion.

For comparison, we also show the corresponding fractions in our field sample as yellow dashed lines. To obtain statistically robust results despite the multiple splits imposed on our galaxy sample (M$_*$, M$_{\text{host}}$, r/r$_{200}$), we make use of the fact that each galaxy in our simulations is `observed' in more than one snapshot, but at different points during its infall into the host group or cluster.  We therefore stack the results for the redshift range $0 \leq z \leq 0.5$ to give a total of $\sim 50\, 000$ data points. We have explicitly verified that using a smaller redshift range has no significant effect other than to increase the statistical uncertainties. 

The first clear influence shown in Fig.~\ref{fig:onset} is that of galaxy stellar mass. More massive galaxies (in the right panels) are overall considerably more likely to contain hot and cold gas and to convert this gas into stars. In the case of hot gas, this comes as no surprise, because more massive galaxies with their deeper potential wells can be expected to accumulate and shock-heat more gas than their lower-mass counterparts (e.g., \citealt{White_Frenk_1991}).  That a larger fraction of massive galaxies have substantial cold gas fractions and are star forming compared to the lowest (stellar) mass galaxies in our sample, however, appears to be at odds with observations, which show that it is in fact {\it low-mass} galaxies that tend to have larger cold gas mass fractions (e.g., \citealt{Dutton_et_al_2011}) and sSFRs (e.g., \citealt{Salim_et_al_2007}). At low masses, this discrepancy is likely a numerical effect: as shown in \citet{McCarthy_et_al_2012b}, the $z=0$ sSFRs in \textsc{gimic} are only numerically well-converged for galaxies with $\log_{10}$(M$_*$ / M$_\odot) \ga 9.7$. This is likely due to the fact that a small number of particles results in a poor sampling of the gas density PDF, keeping in mind that only the highest-density gas forms stars and is classified as `cold' here. On the other hand, the absence of a `mass-quenching' effect \citep{Peng_et_al_2010} in more massive galaxies in our sample is a result of incomplete sub-grid physics (such as the lack of AGN feedback). While this is clearly not desirable and makes it difficult to draw quantitative conclusions, we can still analyse \emph{trends} with environment to gain qualitative insight into the relevant physical processes.  

Leaving aside the dependence on galaxy mass by comparing galaxies with similar M$_*$ (i.e., comparing trends in the same column in Fig.~\ref{fig:onset}), there is also clearly a strong influence of galaxy environment. Within each panel, the fraction of galaxies above the threshold gas mass or star formation rate increases systematically with increasing galaxy distance from the host centre. The behaviour is similar for all three quantities under consideration (hot gas, cold gas and star formation), with hot gas being affected more strongly and out to larger cluster-centric radii. In many cases, the fractions for group and cluster galaxies only approach those for the corresponding field sample at $\sim$ 5 r$_{200}$, and in the case of hot gas and low mass galaxies, there is still a significant discrepancy even at these large radii.

A second influence is that of host mass: almost universally, galaxies around a massive cluster (red lines) are more depleted than galaxies near a low-mass group (black) at the same host-centric distance in units of r$_{200}$, although the difference between different host masses at the same radius is generally smaller than the radial variation. There is also a correlation between the influence of galaxy and host mass: While low-mass galaxies (left-most column) exhibit strong radial variations in both their hot and cold gas content across the whole range of host masses under consideration here, massive galaxies (right-most column) show only a very mild effect on their cold gas content and ability to form stars in environments other than a massive cluster. All of these trends are broadly consistent with what is expected for ram pressure and tidal stripping.

Strictly speaking, the existence of these trends alone does not guarantee that galaxies are actually changing during infall: it is also conceivable that those further away from the host centre were richer in gas and forming stars more actively since they formed. However, in Fig.~\ref{fig:tidal} below we show that the same trends exist in \emph{self-normalised} galaxy properties, such as the ratio between the hot gas mass of a galaxy at a particular radius and at first crossing of 5 r$_{200}$. This is incompatible with a scenario in which the trends presented here are the result of varying galaxy formation conditions: Galaxies are actually changing as they move closer to the host centre.

Finally, we note that there is relatively little difference between binning galaxies according to M$_*$ and M$_\text{host}$ at the snapshot of `observation' (solid lines and bands) and according to the corresponding values at first crossing of 5 r$_{200}$ (M$_*$) and $z = 0$ (M$_\text{host}$) as shown by dashed lines. In the second case, the trends in Fig.~\ref{fig:onset} represent stacked evolutionary tracks of individual galaxies as they move towards the central host region, whereas our default binning method allows galaxies to change bins as a result of either the host mass or galaxy stellar mass varying with time. The `fixed binning' method gives slightly larger fractions of gas-rich and star forming galaxies, as both the host masses and stellar masses generally increase with time. Measuring them earlier than the galaxy `observation' (M$_*$) or later (host mass) therefore on average underpredicts the first while overestimating the second: in both cases galaxies are assigned to bins with stronger environmental influence and therefore appear to be less strongly affected than their counterparts whose stellar mass and host mass were determined at the observation snapshot. In any case, the fact that the difference is small implies that the trends must be predominantly caused by actual changes to the galaxy properties, rather than their bin designations.

\section{The origin of large-scale trends}
We have shown in the previous section that galaxies which are even moderately close to groups and clusters in the \textsc{gimic} simulation are systematically depleted of hot and cold gas, and are consequently less likely to be star-forming, compared to field galaxies of the same stellar mass. In the following, we aim to identify the physical processes responsible for these trends. There are two broad categories of such mechanisms: direct interaction between the host and its galaxies at large radii on the one hand, and indirect effects such as pre-processing of galaxies in groups and/or `overshooting' on the other. We will first address the importance of each of these indirect effects, before investigating the possibility of direct galaxy--host interactions at large radii further below. 

\subsection{Overshooting}

The first possible explanation is that a significant fraction of galaxies in the outskirts of groups and clusters are not actually falling in for the first time. Following their pericentric passage (in most cases well inside r$_{200}$), galaxies on highly elliptical orbits may either permanently escape the host on hyperbolic trajectories or at least reach their apocentre well beyond r$_{200}$ before falling in towards the central region once more. Despite their large distance from the host centre, these galaxies will likely already have experienced strong stripping due to ram pressure and tidal forces during their passage through the inner cluster regions and have thus lost a significant amount of their originally bound gas. As fewer galaxies will return to greater distances from the centre, this effect will naturally lead to a gradual increase in the fraction of star forming, gas rich galaxies with increasing distance beyond the virial radius, without any actual environmental influence at large radii. 

\begin{figure}
\includegraphics[width=\columnwidth]{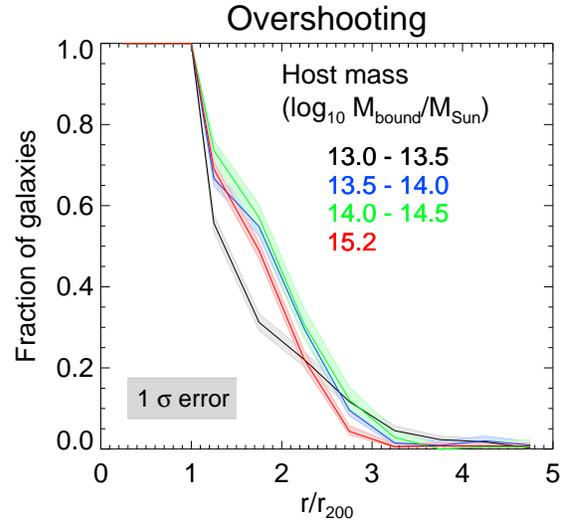}
\caption{Fraction of galaxies which have been within r$_{200} (z)$ of the host centre and therefore, if found at r $>$ r$_{200}$, have already passed their first pericentre. Shaded regions show 1$\sigma$ binomial uncertainties. This is very common within 2 r$_{200}$. Outside 3 r$_{200}$, on the other hand, almost all galaxies are infalling for the first time.}
\label{fig:overshooting}
\end{figure}

The extent of this `overshooting' in our galaxy sample can be judged from Fig.~\ref{fig:overshooting} in which we show the fraction of galaxies that have already ventured into the central r$_{200}$ as a function of host-centric distance\footnote{Note that we take into account that r$_{200}$ evolves with redshift, as the host grows.}. There is no clear dependence on host mass: in all cases, the fraction of overshot galaxies decreases strongly between r$_{200}$ and 3 r$_{200}$. Beyond this radius almost all galaxies are infalling for the first time. The trends in Fig.~\ref{fig:onset} are therefore likely to be substantially affected by overshooting at $r < 3$ r$_{200}$. Beyond this point, however, the overwhelming majority of galaxies are infalling for the first time, which rules out overshooting as an explanation for trends in the far outskirts of groups and clusters. 

\begin{figure}
\includegraphics[width=1.05\columnwidth]{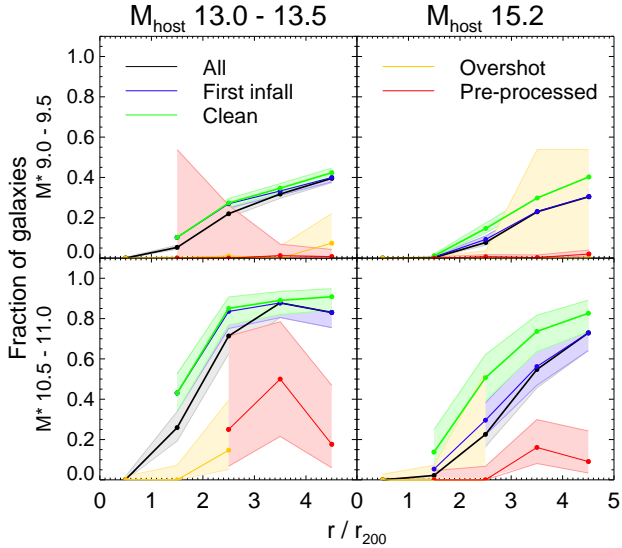}
\caption{Radial trends in the fraction of galaxies with hot gas mass fraction above the threshold of 0.1, using different subsets of our infall galaxy sample. Black lines show all galaxies, as used in Fig.~\ref{fig:onset}, while blue and green lines show only those galaxies that are infalling for the first time and have also never been in a group, respectively. Shaded regions show 1$\sigma$ binomial uncertainties. The green line therefore represents a `clean' sample of galaxies in which any trends are due to direct interaction with the cluster at radii equal to or greater than the current position of the galaxy. Furthermore, yellow and red lines show galaxies that have already travelled through the inner regions of their host cluster or have at some point in the past been satellites in a different subhalo. Where lines do not cover the entire radial range this is due to a lack of galaxies in a given category at certain radii, e.g. overshot massive galaxies in low-mass groups at large radii.}
\label{fig:cleansimple}
\end{figure}

Fig.~\ref{fig:cleansimple} highlights the difference in the radial trends for the hot gas fraction between overshot galaxies and those infalling for the first time; for simplicity and clarity, we show here trends from only two bins each in stellar mass and host mass. While overshot galaxies are almost completely free of hot gas at all host-centric radii (yellow lines), the trend for those galaxies infalling for the first time (blue) is similar to that for the full sample shown in Fig.~\ref{fig:onset}.

\subsection{Pre-processing}
\label{sec:preprocessing}

A second indirect mechanism to explain environmental influence on galaxies at large distances from the host centre is that, in a universe in which structures grow hierarchically, groups and clusters of galaxies are surrounded by other, smaller groups falling into them. Fig.~\ref{fig:onset} showed that strong environmental effects exist near the central regions of even small groups with M $\approx 10^{13} $M$_\odot$ (see also \citet{Balogh_McGee_2010}, \citet{Wetzel_et_al_2012} and \citet{Rasmussen_et_al_2012} who find a similar result in observational data). It is possible that many of the gas-poor, passive galaxies in the outskirts of a big cluster are really sitting in the central region of small groups, which are primarily responsible for the removal of gas. This `pre-processing' effect has been considered as an explanation for reduced star forming fractions beyond r$_{200}$ by a number of authors (e.g., \citealt{Berrier_et_al_2009,McGee_et_al_2009,Lu_et_al_2012,Wetzel_et_al_2012}).

Apart from the obvious case of a galaxy actually being identified as a satellite in an infalling group at the point of observation, there are two other circumstances in which a galaxy can be affected by pre-processing: firstly, a galaxy may \emph{have been} such a satellite in the past, but subsequently ceased to be (for example, because it escaped the group's gravitational attraction or the group itself was tidally disrupted). On the other hand, as explained in Section 2, it is also possible for groups to be accreted onto a massive FoF halo of a galaxy cluster without being disrupted. In this case, the group satellite galaxies become --- formally --- satellites of the cluster instead of the group, which in reality continues to affect them. To identify such `hidden' groups we make use of the fact that, while no longer forming their own FoF halo, they are nevertheless a gravitationally self-bound entity which is detected as a single subhalo by \textsc{subfind}, with the individual galaxies within the hidden group identified as sub-sub-haloes. 

\begin{figure}
\includegraphics[width=\columnwidth]{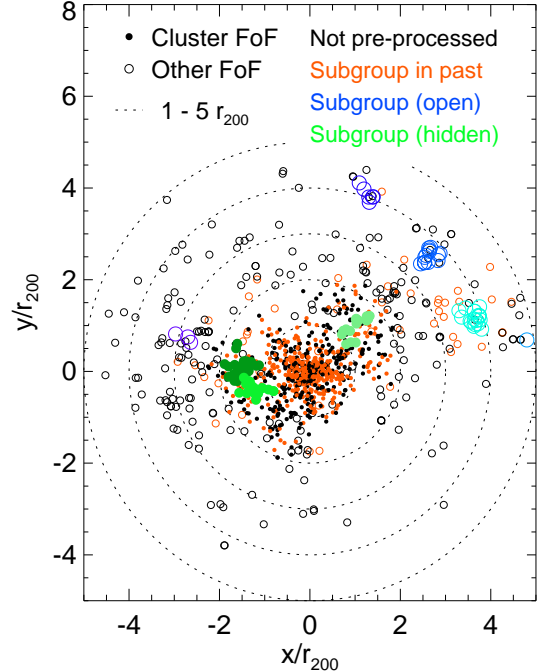}
\caption{Pre-processed galaxies in the massive cluster at z = 0 (coloured circles). Open circles in shades of blue represent galaxies in groups identified as separate FoF halo, while filled circles in green shades show galaxies in `hidden' groups, which are part of the main cluster FoF halo, but form a self-bound sub-halo with total bound mass M$_{\text{bound}} \geq 10^{13} $M$_{\odot}$. Galaxies shown with the same colour belong to the same group. Black circles show cluster galaxies which are not in any group: filled ones are identified as part of the cluster FoF halo, open ones reside in separate haloes. In the same way, orange circles show galaxies that belonged to a group only at z $>$ 0. The dotted circles represent distances of (1, 2, 3, 4, 5) r$_{200}$ from the cluster centre.}
\label{fig:subgroupmap}
\end{figure}

As an illustration, Fig.~\ref{fig:subgroupmap} shows a projected map of all galaxies within 5 r$_{200}$ from the centre of the massive galaxy cluster at redshift z = 0. Galaxies unaffected by pre-processing (i.e., those that have never been satellites in a group with M $\geq 10^{13} \text{M}_\odot$) are shown as black circles, filled ones representing galaxies identified as part of the cluster FoF halo while open ones are not. A handful of `open' groups (not part of the cluster FoF halo) are shown by large open circles in different shades of blue, all of them at relatively large distances from the cluster centre (r/r$_{200} > 3$). Further in, satellites of three hidden groups are shown by large filled circles in shades of green. We note that, looking ahead to Fig.~\ref{fig:filamentmatch}, the location of these hidden groups coincides with strong overdensities of infalling gas (as indicated by their long tails pointing away from the cluster centre). This confirms that these hidden groups are real, physical structures. Furthermore, Fig.~\ref{fig:subgroupmap} also highlights those galaxies which are not member of any infalling group at z = 0, but were in the past, with orange circles (filled/open denoting galaxies within/outside the cluster's FoF halo). A large number of galaxies in the central region (r/r$_{200} \leq 2$) belong to this category, but also a noticeable number of galaxies around the groups at large radii.

\begin{figure}
\includegraphics[width=\columnwidth]{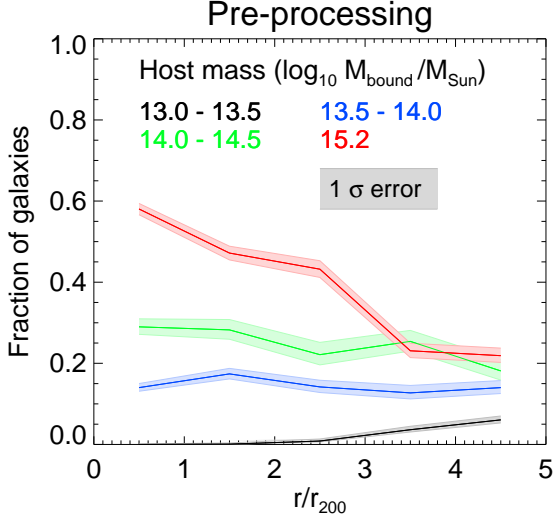}
\caption{Fraction of galaxies which have been satellites of a halo with mass above $10^{13}$ M$_\odot$ other than the host, as a function of galaxy position at time of observation. Different host mass bins are shown in different colours as indicated at the top of the figure. Shaded bands show the corresponding binomial $1\sigma$ uncertainties. Galaxies in more massive hosts are more likely to have been satellites in a group, but with the exception of the most massive cluster there are no clear trends with galaxy position.} 
\label{fig:subgroups}
\end{figure}

Combining all these various types of pre-processed galaxies, Fig.~\ref{fig:subgroups} shows that on average approximately half the galaxies around the massive cluster are affected, increasing towards the centre from $\sim 30$ per cent at 5 r$_{200}$ to $\sim 65$ per cent within r$_{200}$. In lower mass hosts, pre-processing is less common, decreasing to less than 10 per cent in the case of low-mass groups. Even in this case, pre-processing is much more common than overshooting beyond $\sim 3$ r$_{200}$ and is therefore a significant contributor to the radial trends seen in Fig.~\ref{fig:onset}. 

The fraction of pre-processed galaxies which are hot gas rich (M$_\text{gas}/$M$_* > 0.1$) is shown in Fig.~\ref{fig:cleansimple} together with the effect of overshooting discussed above. Pre-processed galaxies are represented by red lines; it is clear that their hot gas fraction is significantly lower than in the full sample (black lines). In the case of low-mass galaxies (top row) the effect is virtually the same as that of over-shooting: in both cases, hardly any affected galaxies have significant amounts of hot gas, irrespective of their distance from the host centre.  The effect is somewhat milder for massive galaxies (bottom row) but still quite significant.

We point out that we only call a galaxy ``pre-processed'' if it was/is a \emph{subhalo} of an infalling group with M$_\text{bound} \geq 10^{13} $M$_{\odot}$, which in practice typically means that it has been within $\sim 1.5$ r$_{200}$ of the group centre. Fig.~\ref{fig:onset} shows that the effect of environment around low-mass groups actually extends out slightly further than this, typically to $\sim$ 2 r$_{200}$. Thus, we have adopted a `strong' definition of pre-processing in the present work. This choice is justified by the trends seen in Fig.~\ref{fig:cleansimple}: Galaxies which meet our definition of pre-processing differ markedly in their properties such as hot gas content from those which do not. We have experimented with `weaker' definitions, such as lower mass thresholds and larger radii around group centres and found that they lead to less pronounced difference between `pre-processed' and `not pre-processed' galaxies, simply because not all galaxies included under a `weak' definition are actually affected by their proximity to a sub-group.

\begin{figure*}
\includegraphics[width=2\columnwidth]{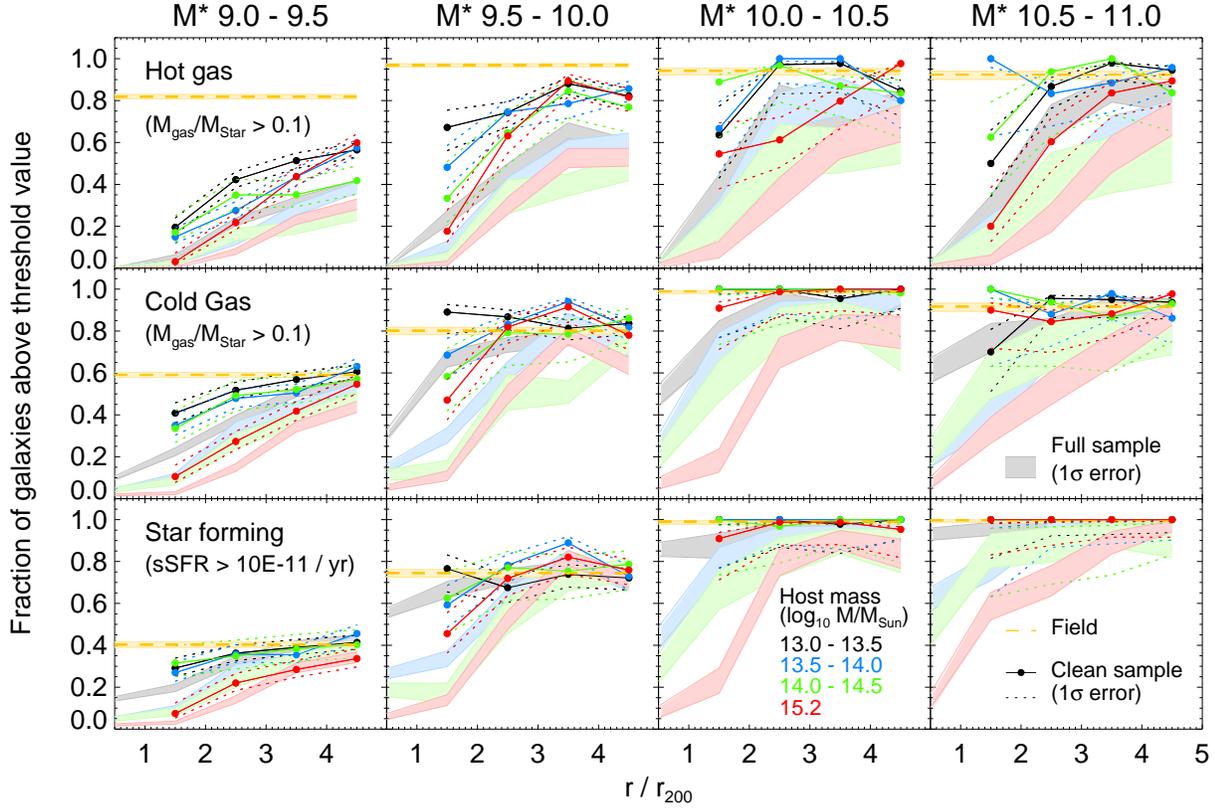}
\caption{Comparison between radial trends for the `clean' galaxy sample, containing only galaxies which have never been within r$_{200}$, and have never been satellites in another halo than the main host (shown by solid lines, dotted lines give statistical $1\sigma$ uncertainties) and the full sample from Fig.~\ref{fig:onset} (shaded regions showing statistical $1\sigma$ binomial uncertainties, for clarity we have omitted the main trend for this sample). In the clean galaxy sample, radial trends are significantly weaker and start at smaller radii, with the exception of hot gas and low-mass galaxies.}
\label{fig:cleanfull}
\end{figure*}

Combining the effects of overshooting and pre-processing, we can form a `clean' sample of galaxies affected by neither: these are infalling for the first time directly from the field. In Fig.~\ref{fig:cleansimple} they are shown in green; unsurprisingly their hot gas fractions are higher than in any of the four other samples, at all radii. A detailed comparison between the full and clean galaxy samples, for all M$_*$ and M$_\text{host}$ is shown in Fig.~\ref{fig:cleanfull}.  The shaded bands represent the full sample (identical to Fig.~\ref{fig:onset}), whereas the solid and dotted lines show the corresponding fraction of galaxies in the clean sample and their statistical uncertainties, respectively. 

In the case of cold gas and star formation, the trends are much weaker in the clean sample, and for massive galaxies with M$_*$ $\ga 10^{10} \text{M}_\odot$ there is virtually no trend remaining - i.e., at all radii the fraction of galaxies with appreciable cold gas or star formation is similar to that in the field. Where trends exist (for low-mass galaxies), they are strongest in the case of galaxies near the massive cluster.  However, a much stronger environmental influence remains in the case of hot gas, with galaxies of all masses that we have explored being affected out to large radii from the centres of both group and cluster hosts.

We had pointed out in the previous section that the fraction of hot gas rich low mass galaxies near groups and clusters is still significantly lower than in the field even at $\sim 5$ r$_{200}$. This discrepancy is smaller in the clean than in the full sample, but still significant, and turns out to be a consequence of our `strong' pre-processing definition as discussed above. The reason is that our field galaxies are by definition all centrals far away from groups and clusters, whereas even our clean sample of galaxies near groups and clusters still contains satellites in low-mass haloes (equivalent, for example, to the Milky Way satellites) and/or those near a sub-group, but outside its friends-of-friends halo. We have tested this by adopting a `very weak' pre-processing definition, where we impose no lower mass threshold at all, combined with a 5 r$_{200}$ radius of influence around sub-groups with M$_\text{bound}$ $> 10^{13}$ M$_\odot$. In the complementary `very clean' sample (galaxies on first infall which are not pre-processed even under this definition) the fraction of hot gas rich galaxies still increases with radius out to $\sim$ 5 r$_{200}$, but slightly stronger than in the `clean' sample shown in Fig.~\ref{fig:cleanfull}, and therefore agrees with that in the field sample at this radius. 

We finally note that, due to the large number of galaxy groups in the cosmologically rare $+2\sigma$ \textsc{gimic} sphere, our sample includes more groups which are located relatively close to a more massive halo than a representative cosmological volume. To ensure that this does not bias our results, we have verified that the trends shown in Figs.~\ref{fig:onset} and \ref{fig:cleanfull} do not change significantly if we only include `isolated' groups far away from any more massive haloes. 

\subsubsection{Summary}

Our conclusions so far from this section may be summarised as follows: the strong radial trends in cold gas and star forming fraction seen in Fig.~\ref{fig:onset} are largely caused by overshooting and pre-processing, especially in the case of massive galaxies with M$_*$ $\ga 10^{10}$ M$_\odot$, as seen in Fig.~\ref{fig:cleanfull}. Out of these two indirect mechanisms, overshooting is generally dominant within $\sim 2$ r$_{200}$, while at larger radii the trends are mostly due to pre-processing.  The lowest-mass galaxies, on the other hand, show appreciable radial trends in their retention of cold gas and star forming activity even when pre-processing and overshooting are excluded.  In terms of hot gas, the radial variations for both low- and high-mass galaxies are very similar in the full and `clean' samples. Therefore, a direct influence of the group or cluster environment must extend out to at least $\sim$ 5 r$_{200}$. In the remainder of the paper, we investigate this influence, and therefore use only the clean galaxy sample from here on\footnote{Although this does include some `weakly pre-processed' galaxies, as discussed above, we have verified that none of our results in the following sections are changed significantly when using the -- much smaller -- `very clean' sample instead.}

\subsection{Direct galaxy--host interaction: tidal and ram pressure stripping}

In the previous section, we showed that neither pre-processing nor overshooting can can fully account for the depletion of cold gas and quenching of star formation in low-mass galaxies found in the outskirts of groups and clusters.  Nor can it account for the removal of hot gas of both low-mass and high-mass galaxies.
This means that a process must exist by which galaxies can be influenced {\it directly} by the hosts at large distances from their centre. In this section, we argue that this process is most likely direct ram pressure stripping due to interaction of the galaxies with an extended hot gas halo surrounding groups and clusters. 

\begin{figure}
\includegraphics[width=\columnwidth]{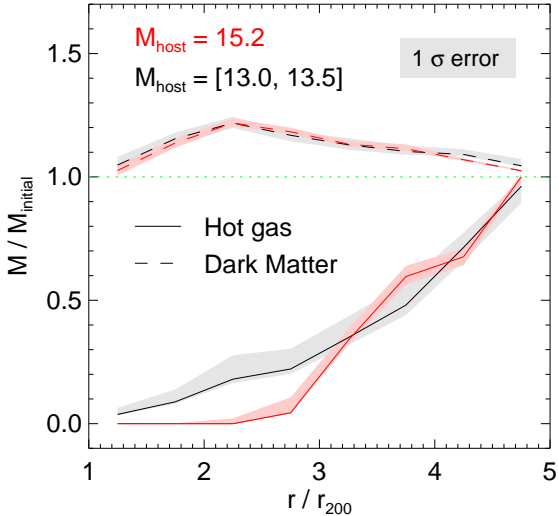}
\caption{Evolution of the hot gas and dark matter content of low-mass galaxies falling into hosts for the first time without having been affected by pre-processing. Solid lines show the median hot gas mass, dashed lines the median dark matter mass, both normalised for each galaxy to the respective values at first crossing of 5 r$_{200}$. The red lines represent galaxies falling into the massive cluster in the +2$\sigma$ simulation, black ones those falling into low-mass groups. Shaded regions show 1$\sigma$ confidence intervals as described in the text. There is a clear difference between the evolution of the hot gas content, which decreases within $\sim$ 5 r$_{200}$, and dark matter, which, irrespective of halo mass, \emph{increases} slightly until $\sim$ 2 r$_{200}$. This implies that at large distances from the host centre, the hot gas is removed by a process targeting exclusively baryons, such as ram-pressure stripping, and not by a more indiscriminate one such as tidal stripping.}
\label{fig:tidal}
\end{figure}

Besides ram pressure, there are several other mechanisms which could also be responsible for removing gas from galaxies, in particular tidal stripping due to the cluster potential or galaxy--galaxy interactions. In contrast to ram pressure, these processes can be expected to affect not only the (hot) gas content of galaxies, but also the similarly extended dark matter haloes. In Fig.~\ref{fig:tidal} we show the evolution of the hot gas and dark matter content of `clean' galaxies falling into both the massive cluster and low-mass groups, focusing on low-mass (log M$_*$/M$_\odot$ = [9.0, 9.5]) galaxies for which the environmental effect is strongest (see Fig.~\ref{fig:cleanfull}), although we have verified that similar trends are also seen in more massive galaxies. Making use of our galaxy tracing results, we normalise the hot gas and dark matter masses by their respective values at first crossing of 5 r$_{200}$. Any deviation from unity in these `self-normalised' values is necessarily the result of changes occuring within individual galaxies, and not due to potentially differing galaxy formation conditions at different distances from the host centre. Shaded bands in Fig.~\ref{fig:tidal} represent the 1$\sigma$ confidence intervals, obtained by scaling the intrinsic scatter in the distribution of M/M$_\text{initial}$ by the square root of the number of individual galaxies in each bin. 

Hot gas and dark matter clearly evolve very differently: from 5 r$_{200}$ onwards, the hot gas mass decreases steadily with decreasing radius (solid lines) with an overall stronger effect in the case of the cluster than low-mass groups (red and black lines, respectively).  In the former, the majority of galaxies have lost all of their hot gas around 3 r$_{200}$ and even in low-mass groups the median hot gas mass of galaxies is reduced to $\sim 20$ per cent by this point, compared to the value at 5 r$_{200}$.  The mass of the dark matter halo, on the other hand, remains nearly constant (dashed lines in Fig.~\ref{fig:tidal}) at these large radii, independent of host mass, and in fact increases slightly until $\sim$ 2 r$_{200}$ due to continuing accretion of dark matter. This implies that the removal of hot gas in the group and cluster outskirts is due to a process targeting exclusively baryons while leaving the dark matter halo basically untouched --- precisely the behaviour that would be expected from ram pressure stripping.

\subsubsection{Expected effect of ram pressure}

\begin{figure*}
\includegraphics[width=2\columnwidth]{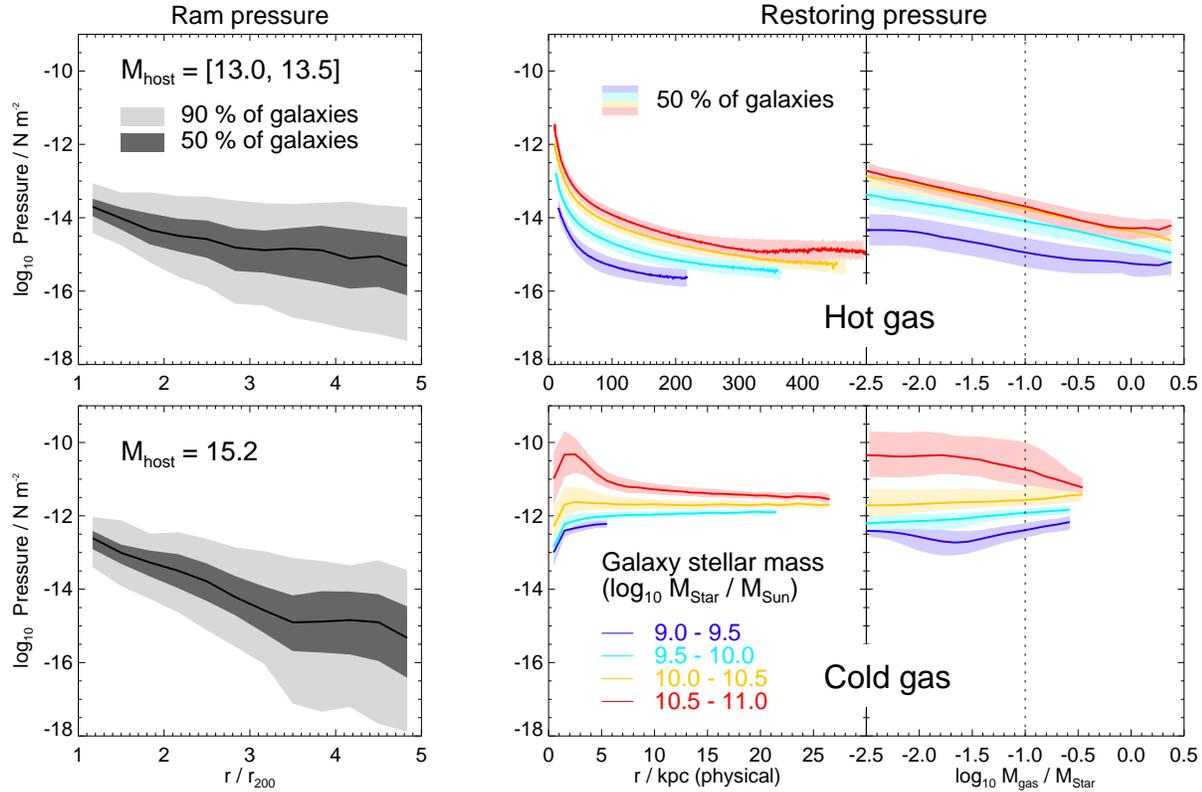}
\caption{Comparison of ram and restoring pressure on galactic gas. For ram pressure (left), the top panel shows low-mass groups, the bottom panel a massive cluster. The solid black line shows the median trend, dark and light grey bands enclose 50 and 90 per cent of galaxies. For restoring pressure (middle/right columns), the top panels show hot gas whereas the bottom panels show cold gas; in both cases differently coloured lines show galaxies of different stellar mass. In the middle column, the restoring pressure is plotted as a radial profile, in the right as a function of enclosed specific gas mass. See text for details. This comparison predicts substantial stripping of hot gas in hosts of all masses. Cold gas, on the other hand, is only expected to be significantly stripped in the central cluster regions. This agrees well with what is observed in the \textsc{gimic} galaxies.}
\label{fig:pramprest}
\end{figure*}

To test the ram pressure stripping hypothesis further, we directly compare the ram pressure and gravitational restoring forces on galactic hot and cold gas in Fig.~\ref{fig:pramprest}. For each galaxy, ram pressure is computed as
\begin{equation}
  P_{\text{ram}} = v^2_\text{ICM}\,\, \rho_\text{ICM}
\end{equation}
where $v_\text{ICM}$ is the velocity of the galaxy relative to the surrounding ICM of density $\rho_{\text{ICM}}$.  To determine these two values, we select for each galaxy the N = 3000 closest gas particles\footnote{We have experimented with various values of N and found N = 3000 to be the optimal value. For too low N, particle-to-particle scatter in velocity and density becomes noticeable, whereas at too high values of N we are no longer determining the local properties.} which are not members of any gravitationally bound subhalo (except for the main subhalo in a host group or cluster). This ensures that our measurements of ICM density and velocity are not influenced by particles in nearby galaxies, but actually represent the ICM. To exclude contamination by gas accreted by, or stripped from, the galaxy under consideration, we also exclude any gas particles that have previously been, or will subsequently be, bound to it. The relative velocity $v_\text{ICM}$ is then simply the mass-weighted average velocity of these 3000 particles in the galaxy rest frame. 

The left column of Fig.~\ref{fig:pramprest} shows, for `clean' (not pre-processed or overshot) galaxies in low-mass groups (top) and the massive cluster (bottom), the distribution of resulting ram pressure values with varying distance from the host centre. The median trend is given by the thick black line, while the dark and light grey regions enclose 50 and 90 per cent of all galaxies, respectively. In both low-mass groups and the massive cluster, ram pressure is increasing towards the centre, but the trend is stronger in the latter case where it varies by approximately 3 orders of magnitude between 5 r$_{200}$ and r$_{200}$ as opposed to `only' 2 orders of magnitude in low-mass groups over the same radial range. While the ram pressure experienced by galaxies in the outskirts of both groups and clusters is similar (at same r/r$_{200}$), galaxies near the \emph{centre} of a cluster therefore experience considerably higher ram pressure levels than their group counterparts. We show below that this is primarily a consequence of the higher orbital velocities of galaxies in massive clusters.  Apart from this overall trend, there is also substantial scatter in the ram pressure values, in particular in the outer regions. Galaxies at a distance of 4 -- 5 r$_{200}$ from the centre of a massive cluster can experience ram pressure differing by about 5 orders of magnitude, a range considerably larger than the systematic variation with radial distance. We will investigate the origin and implications of this scatter in Section \ref{sec:filaments} below.

Gas will be removed from the infalling galaxies if the ram pressure exceeds the gravitational restoring pressure (restoring force per unit area) exerted by the galaxy. Following \citet{McCarthy_et_al_2008}, we compute this quantity as

\begin{equation}
  P_{\text{restore}} (r) = \frac{\alpha G M(<r) \rho(r)}{r}
  \label{eq:pramprest}
\end{equation}
where $M(<r)$ is the total mass within galacto-centric radius $r$, $\rho(r)$ the density of the gas phase (hot or cold, defined as discussed in section \ref{sec:onset}) under consideration, and $\alpha$ is a geometric constant of order unity. \citet{McCarthy_et_al_2008} find $\alpha = 2$, which we adopt for our calculations as well, although the exact choice of this parameter has no influence on our conclusions. Using equation \eqref{eq:pramprest}, we compute hot and cold gas restoring pressure profiles for all \textsc{gimic} field galaxies (centrals that have never been within 5 r$_{200}$ of a group or cluster) as these represent the `initial condition' of galaxies before infall into a group or cluster. To connect the pressure comparison directly to the gas content, we furthermore find for each field galaxy in our sample the restoring pressure at the radius enclosing a series of specific hot and cold gas masses (i.e., M$_\text{gas}$/M$_*$) in the range $-2.5 \leq \text{log}_{10} \text{M}_\text{gas}/\text{M}_* \leq 0.5$. The resulting trends, median-stacked in bins of similar stellar mass are shown in the middle and right columns of Fig.~\ref{fig:pramprest} and give the typical level of ram pressure required to strip a galaxy to a given gas mass or outer limiting radius, respectively. We note that, in stacking galaxies, we include at each point only those that actually have hot or cold gas of this mass or extending out to this radius and only show those data points where this is the case for at least 5 per cent of the galaxies to give a meaningful picture of how tightly bound the gas typically is. For hot gas, we show the restoring pressure profiles over a radial range from 0 to 500 kpc (physical), but because the cold gas component is much more centrally concentrated, we use a smaller radial range from 0 to 30 kpc here.

It is evident that cold gas is much more tightly bound than hot gas when comparing the middle/right top and bottom panels. The typical restoring pressure on cold gas ranges from 10$^{-11}$ Pa in the most massive galaxies (red) to 10$^{-13}$ Pa in low-mass galaxies (blue) at the radius enclosing 0.1 M$_*$ in cold gas. By comparison, even in massive galaxies, the corresponding restoring pressure on hot gas is only 10$^{-14}$ Pa, which drops to 10$^{-15}$ Pa in low-mass galaxies. The reason for this is that cold gas is not only denser, but also sits much closer to the galactic centre. 

By comparison, the typical \emph{ram pressure} reaches a maximum level of 10$^{-12.5}$ Pa, in the case of the massive cluster near r$_{200}$, with some galaxies reaching levels up to 10$^{-12}$ Pa. This is clearly too low to strip cold gas in massive galaxies, but just sufficient for those with lower stellar masses. Outside $\sim$ 2 r$_{200}$ and in less massive hosts, however, no galaxies experience sufficient ram pressure to directly strip cold gas. Hot gas on the other hand, bound by approximately two orders of magnitude less tightly, can be stripped efficiently: even massive galaxies (red) can be affected out to $\sim$ 2 -- 3 r$_{200}$ in clusters, and many low-mass galaxies are subject to sufficient ram pressure ($\sim 10^{-15}$ Pa) even at 5 r$_{200}$. Even in low-mass groups, hot gas can be expected to be stripped out to $\sim$ 2 r$_{200}$ in all galaxies, and in a large fraction ($> 25$ per cent) out to 5 r$_{200}$. 

These expectations agree well with the actual evolution of the gas content as seen in Fig.~\ref{fig:cleanfull}. In particular, stripping of hot gas of galaxies (top) is seen out as far as 5 r$_{200}$ in all environments, whereas cold gas is only affected in low-mass cluster galaxies, as predicted from our pressure comparison. 

We finally note that the restoring pressure profiles themselves show an interesting difference between cold and hot gas: those for cold gas are relatively flat outside $\sim$ 5 kpc while the hot gas profiles show a steady decline from the central region outwards.  This suggests that hot gas is stripped gradually from the outside as the ram pressure acting on a galaxy increases, whereas when the cold gas finally begins to be stripped, virtually all of it will be removed over a short time scale.

\section{Influence of filaments}
\label{sec:filaments}
\subsection{Origin of the ram pressure scatter}

While Fig.~\ref{fig:pramprest} confirms that there is a general trend to higher ram pressure values towards the host centre, it also reveals strong scatter, particularly in the outer regions. At r $\sim$ 5 r$_{200}$ from the centre of the big cluster, the ram pressure varies between galaxies at the \emph{same} distance from the centre by five orders of magnitude, substantially more than the variation in the median ram pressure over the radial range considered here. In this section, we investigate the origin of this scatter and its implications. 

An obvious possibility is that we have so far only distinguished between galaxies by their \emph{radial} distance from the host centre, thereby implicitly assuming that our hosts are spherically symmetric systems. This is rather unlikely: it is a long-standing prediction of cosmological simulations that groups and clusters of galaxies are triaxial systems linked by filaments of both dark matter and gas, and there is now increasing observational evidence that this is indeed the case (e.g., \citealt{Dietrich_et_al_2012}; \citealt{Planck_et_al_2012}). Galaxies falling in along these filaments have a very different infall experience from those accreted through largely empty regions (voids), which we now investigate in detail. As before, we focus exclusively on the `clean' galaxy sample which are infalling for the first time without having been affected by pre-processing.

\begin{figure}
\includegraphics[width=\columnwidth]{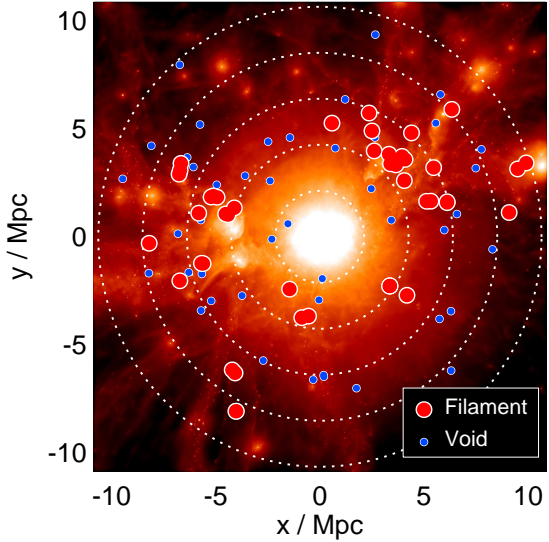}
\caption{Gas filaments around the massive cluster at $z = 0$ (background image, lighter colours corresponding to higher gas surface density). There is a good correlation between dense regions and the location of galaxies that we identify as in filaments (red points). Void galaxies (blue points) show no such strong clustering. The dotted white rings mark projected radii of 1 - 5 r$_{200}$.}
\label{fig:filamentmatch}
\end{figure}

As a parameter to distinguish between galaxies in filaments and voids, we choose the `local overdensity', which we define as 
\begin{equation}
  \Delta_\rho = \rho_{\text{local}} / \rho_{\text{profile}} (r_{\text{galaxy}})
  \label{eq:rholocal}
\end{equation}
where $\rho_\text{local}$ is the locally determined ICM density obtained as discussed in the previous section and $\rho_\text{profile}(r_\text{galaxy})$ is the corresponding spherically averaged gas density at the radius of the galaxy. In a perfectly spherically symmetric host, each galaxy would have a value of $\Delta_\rho = 1$; in a more realistic system, galaxies in filaments are those with the highest $\Delta_\rho$ while those in voids have the lowest values. Having determined $\Delta_\rho$, we then bin galaxies in each snapshot according to host mass and rank the galaxies in each bin by $\Delta_\rho$. The highest quartile in each bin is identified as filament galaxies, and the lowest as those in voids. This ensures that we have equal numbers of filament and void galaxy data points and that both also have the same distribution in redshift and host mass. In Fig.~\ref{fig:filamentmatch} we show the locations of both filament and void galaxies in the massive cluster at z = 0, superimposed on a map of the ICM gas density. As expected, filament galaxies are strongly spatially clustered and are mostly found in two bands fanning out towards the top right and the left, which is exactly the region where prominent filaments can be seen in the gas density map. The void galaxies, however, are much more evenly spread and tend to avoid the filament regions.\footnote{Note that this is a projected map, so that some blue points (void galaxies) may still \emph{appear} to lie in regions of high gas density.} We take this as confirmation that our classification based on $\Delta_\rho$ is a physically meaningful distinction between void and filament galaxies.   

\begin{figure*}
\includegraphics[width=2\columnwidth]{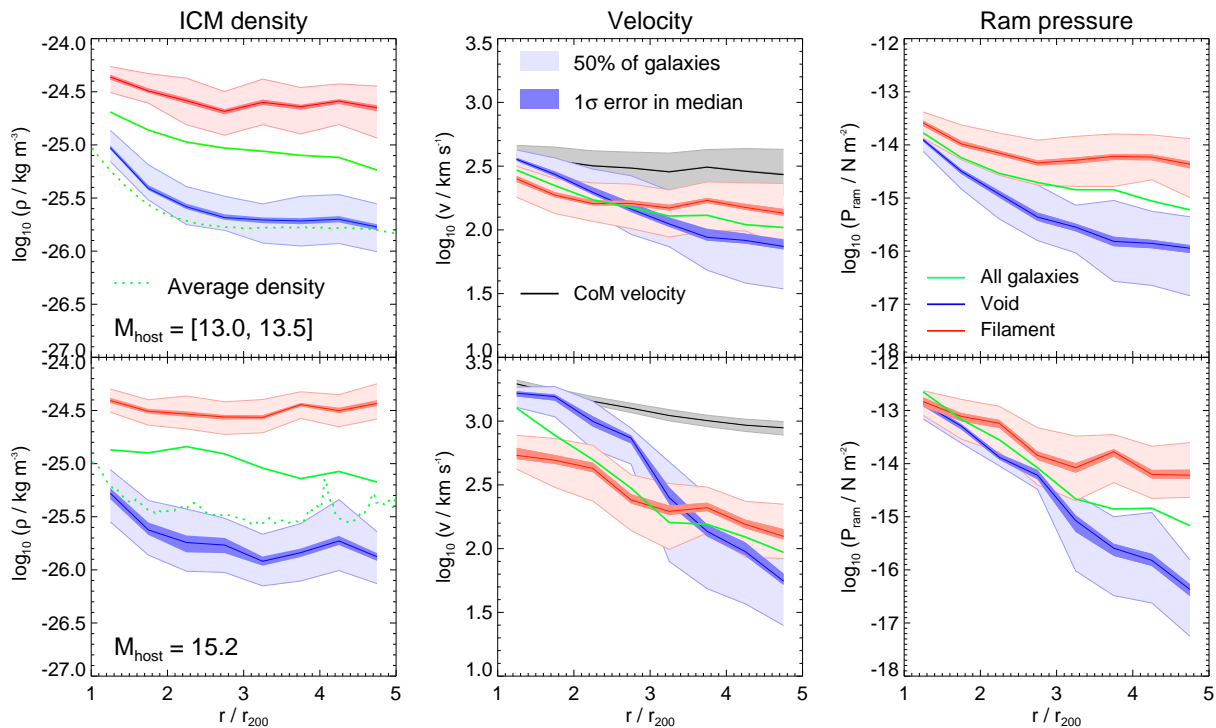}
\caption{Density (\emph{left column}), relative velocity (\emph{middle column}) and resulting ram pressure (\emph{right column}) of the ICM around `clean' galaxies. Low-mass groups are shown at the top, massive clusters in the bottom row, each split into filament galaxies (red) and those in voids (blue) as described in the text; green lines represent the undivided sample. Also shown is the stacked spherical density profile (green dotted) and the velocity with respect to the host centre of mass frame (black). The median and 50 per cent spread of each bin are shown by solid lines and shaded regions respectively; the darker shaded region around the median line gives the 1$\sigma$ confidence interval on the median (which is small due to the large sample size). The radial trends in ram pressure are mostly due to radial variation of velocity, whereas the scatter is dominated by variation in local density.}
\label{fig:rampressureanalysis}
\end{figure*}

As a first step, we show in Fig.~\ref{fig:rampressureanalysis} a decomposition of the ram pressure values from Fig.~\ref{fig:pramprest} (right column) into ICM density (left) and its velocity relative to the galaxy (middle). Galaxies near low-mass groups are shown on top and those around the massive cluster in the bottom row. In both cases, the median trends for filament galaxies are represented by red lines, void galaxies in blue, while the green solid line shows the overall trend (irrespective of $\Delta_\rho$). In the left column showing ICM density, we furthermore show the median-stacked spherical density profile as a dotted green line; in the middle column the velocity of the galaxy relative to the host centre-of-mass frame is shown in black. 

The difference in ICM density between filament and void galaxies is evidently quite strong with the latter around an order of magnitude lower. The difference is least pronounced near r$_{200}$ (0.6 and 0.9 dex in groups and the cluster, respectively), and increases outwards, reaching 1.1 and 1.5 dex at 5 r$_{200}$. In both cases, the density trends are relatively flat, in particular in the case of filament cluster galaxies where there is no significant radial variation outside r$_{200}$. The increased density difference between voids and filaments towards larger radii is driven mostly by a moderate decrease in the density around void galaxies. Including scatter within each band, ICM densities around 5 r$_{200}$ vary by approximately 1.5 orders of magnitude in groups and two in clusters, thus accounting for about half of the overall ram pressure scatter. As expected, the typical ICM density for all galaxies (green) lies approximately halfway between the filament and void trends. It is worth pointing out that the \emph{overall} spherical density profile (green dotted) is generally lower than the ICM near galaxies, indicating that galaxies preferentially live in overdense regions. This is particularly true in low-mass groups, where the spherical density is even lower than the median density around void galaxies --- in this environment, `voids' really represent the average density, whereas filaments are highly overdense regions. In clusters, on the other hand, voids are actually underdense with respect to the spherical average.

The ICM velocity behaves in a rather different way. While the overall radial variation in low-mass groups (middle top panel, green line) is comparable to that in density, it is much stronger in the case of the cluster (middle bottom panel), where it varies by significantly more than an order or magnitude, decreasing with increasing radial distance. The velocity between a galaxy and its surrounding ICM is almost always considerably lower than that with respect to the host centre-of-mass frame (black), the difference being greatest at largest radii (more than a factor of 10 at its most extreme, $\sim$ 1000 and 100 km/s, respectively). The closest match between both is found for void galaxies near r$_{200}$ where the two velocities virtually agree, indicating that in this case the ICM is largely stationary in the host frame, whereas it is infalling together with the galaxies at larger radii. Perhaps surprisingly, the radial trends are \emph{stronger} for void galaxies, so much so that at 5 r$_{200}$, their velocity relative to the ICM is actually lower than in filament galaxies at the same radius outside $\sim$ 3 r$_{200}$. But even within this radius, the much higher ICM density around filament galaxies means that they experience \emph{higher} levels of ram pressure (see right column) throughout their infall than void galaxies, with no evidence of `dynamical shielding' by which the co-flow of filament gas could protect galaxies from ram pressure by the surrounding gas (except possibly very close to r$_{200}$, see below).

As the velocity enters into ram pressure in quadrature, its strong radial trend is the main cause for the radial variation in ram pressure (accounting for a variation of $\sim$ 1 order of magnitude in P$_\text{ram}$ for filament galaxies and three orders for those in voids). Looking at the right-hand column, almost the entire extent of \emph{radial variation} in ram pressure can therefore be attributed to an increased velocity nearer the centre, while its \emph{scatter} is strongly influenced by variation in ICM density at the same radius. We next explore the implications of this on the properties of void and filament galaxies.

\subsection{Effect on gas fractions}
The results above show that galaxies in filaments will, on average, experience higher levels of ram pressure than those in voids. However, this does not necessarily imply that these galaxies are also less likely to contain significant amounts of gas: it is plausible, for example, that galaxies falling in along a filament could accrete more gas prior to stripping, since far away from the host centre gas is evidently much more abundant within filaments than outside (see Fig.~\ref{fig:filamentmatch}). 

\begin{figure*}
\includegraphics[width=2\columnwidth]{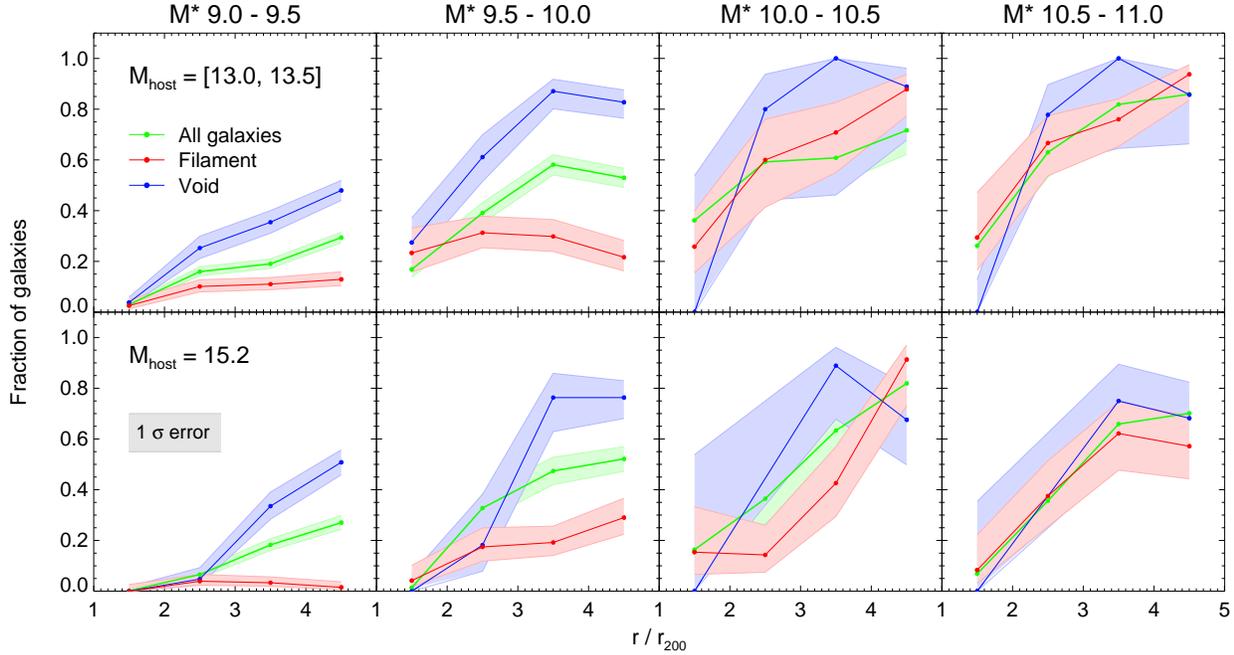}
\caption{Radial trends in the fraction of hot gas rich (log$_{10}$ M$_\text{hot gas}/\text{M}_* \geq -0.5$) `clean' galaxies of varying stellar mass (increasing left to right) in small groups (top) and the massive cluster (bottom). Red lines represent filament galaxies, blue those in voids, and green all galaxies. For low-mass galaxies in particular, galaxies in filaments are significantly less likely to have a high hot gas content than those falling in through voids. For more massive galaxies, however, there is no significant difference between the two infall routes.}
\label{fig:angularvariation}
\end{figure*}

To see which influence is stronger, we plot in Fig.~\ref{fig:angularvariation} the fraction of galaxies with substantial hot gas (log$_{10}$ M$_\text{hot gas}$/M$_* \geq -0.5$) as a function of distance from the host centre.\footnote{This threshold is a factor of $\sim$ 3 higher than that which we have used above. While both limits give rise to similar overall trends, we have found that this threshold value leads to the clearest separation between void and filament galaxies.} Again, we distinguish between galaxies in filaments and voids (red and blue respectively, with green lines showing the overall sample), for both low mass groups (top) and the massive cluster (bottom). The solid lines show the fractions themselves, while the shaded bands represent the statistical 1$\sigma$ binomial uncertainties, as described in section \ref{sec:onset}.  As we are now once more looking at internal galaxy properties, we split our sample by stellar mass, each bin represented in a different column.

The clearest trends are visible for low mass galaxies with log$_{10}$ (M$_*$/M$_\odot) \leq 10.0$. In both the big cluster and low-mass groups, they reveal a clear influence of local environment, in particular in the outer regions (beyond $\sim$ 3 r$_{200}$), in the sense that galaxies in filaments (red) are much less likely to be hot gas rich than those in voids (blue): even at a distance of 5 r$_{200}$ from the cluster centre, only $\sim 5$ per cent of filament galaxies in the lowest stellar mass bin are hot gas rich, compared to $\sim 50$ per cent of void galaxies (bottom left panel). Evidently, the increased level of ram pressure stripping experienced in filaments is more than enough to negate any potential benefit in gas accretion these galaxies might have experienced. The difference is slightly smaller in low-mass groups, but even here a high hot gas content is found in only $\sim 10$ per cent of filament galaxies as opposed to $\sim 50$ per cent in voids. In the case of low-mass galaxies, the influence of filaments is therefore clearly an \emph{increased stripping} of hot gas at large radii. 

Note that like all results presented in this paper, these trends are found at low redshift ($z \leq 0.5$). We have verified that at high redshift ($z = 2$), an equally large fraction of our filament and void galaxies are hot gas rich. The clear discrepancy even at 5 r$_{200}$ between the two galaxy populations at low redshift therefore indicates that the stripping influence of filaments extends even beyond 5 r$_{200}$, and also implies that this effect is predominantly a low-redshift phenomenon. We intend to pursue this result in a future study, together with the general redshift evolution of environmental trends which is now increasingly probed observationally (e.g., \citealt{Balogh_et_al_2011,McGee_et_al_2011,Snyder_et_al_2012}).

For galaxies of higher stellar mass, the smaller sample size makes it more difficult to draw statistically robust conclusions as to the difference between filament and void galaxies. Any potential difference is certainly smaller than in the lower stellar mass bins --- within the large statistical uncertainties, almost all trends are consistent with no difference at all between filament and void galaxies. It is particularly unfortunate in this respect that there are very few high-mass void galaxies: in the same way as galaxies in general are preferentially found in overdense regions (see above), the filaments are clearly a more attractive habitat for massive galaxies than voids, as is evident from the different magnitude of the errors. There may be a hint of evidence for an \emph{increased} hot gas fraction in filament galaxies near r$_{200}$, in particular in the massive cluster (bottom row). Keeping in mind that the ram pressure levels between filament and void galaxies are most closely matched at small radii, this might therefore be an indication for a \emph{small} amount of dynamical shielding by filaments once galaxies approach r$_{200}$. In any case, this difference is small compared to the very strong influence of filaments in the outer regions.

\section{Summary and Discussion}

We have analysed the gas content and star formation rate of galaxies in the vicinity of groups and clusters in the \textsc{gimic} suite of hydrodynamic cosmological simulations. By using zoomed initial conditions, these simulations include rare objects such as a massive galaxy cluster while simulateneously resolving individual galaxies.  We have focussed on determining how far out from the host centre galaxies are environmentally affected, as well as finding the physical mechanisms behind this influence. Our main conclusions may be summarised as follows:

\begin{enumerate}

\item The fraction of galaxies with large hot and cold gas fractions (M$_\text{gas}$/M$_* > 0.1$) decreases systematically with decreasing distance from the host centre; the same is true for the fraction of star forming galaxies (sSFR $ > 10^{-11}$ yr$^{-1}$). These trends extend out to very large distances from the host centre (up to $\sim$ 5 r$_{200}$, corresponding to $\sim 10$ Mpc for a massive cluster). 

\item In terms of cold gas and star formation, the radial trends are explained by galaxies past first pericentre (`overshooting', especially within $\sim 2 - 3$ r$_{200}$) and those having been `pre-processed' in  infalling groups (up to $\sim$ 60 per cent of galaxies, increasing with increasing halo mass.  However, in the case of low-mass galaxies ($9.0 \leq \text{log}_{10} (\text{M}_*/\text{M}_\odot) \leq 9.5$), radial trends extending out to 5 r$_{200}$ are still seen even among those which are not pre-processed and are infalling for the first time.

\item In the case of hot gas, the radial trends cannot be explained solely by overshooting and/or pre-processing. This implies a direct interaction with the host group/cluster out to very large radii.

\item We have shown that the ram pressure exerted by the ICM is strong enough to significantly strip the hot gas haloes around both low- and high-mass galaxies out to several multiples of r$_{200}$.  However, it is not strong enough to significantly strip cold gas far beyond r$_{200}$ (except for low-mass galaxies with M$_* < 10^{9.5}$ M$_\odot$ in a massive cluster). This agrees well with the actual gas content of \textsc{gimic} galaxies.

\item At large distances from the host centre (r $ \ga 2$ r$_{200}$) there is a substantial difference between the velocity of galaxies relative to the host centre-of-mass frame and relative to their surrounding ICM (the latter being lower), increasing with host mass and distance from the host centre. For low-mass groups (M$_\text{host} \sim 10^{13}$ M$_\odot$) the difference is a factor of $\sim 3$, increasing to $\sim 10$ for the massive cluster (M$_\text{host} \sim 10^{15}$ M$_\odot$) simulated in \textsc{gimic}. This implies that calculations based on the centre-of-mass velocity will significantly overestimate the ram pressure exerted on galaxies at large r/r$_{200}$.

\item At large distances from the host centre, there is very large scatter in the ram pressure values experienced by different galaxies at similar r/r$_{200}$, reaching a factor of $\sim 10^5$ in the case of the massive cluster. This scatter is dominated by variations in ICM density: in filaments, ram pressure can be up to two orders of magnitude \emph{larger} than in the lowest-density regions. 

\item Because they experience higher ram pressure, low-mass galaxies (log$_{10}$ M$_*/\text{M}_\odot \la 9.5$) in filaments are significantly less likely to contain substantial amounts of hot gas (log$_{10}$ M$_\text{hot gas}/\text{M}_* \geq -0.5$) than those infalling through voids at the same distance from the host centre ($\sim 5$ vs.~$\sim 50$ per cent at a distance of 4 -- 5 r$_{200}$ from the massive cluster for the lowest mass galaxies in our sample). The difference is less pronounced for more massive galaxies.

\end{enumerate}

Our result that the properties of galaxies near groups and clusters but outside r$_{200}$ differ from those at much greater distances agrees with observational studies, which are increasingly supporting a picture in which the influence of environment reaches well beyond this radius (e.g., \citealt{Balogh_et_al_1999}; \citealt{Lu_et_al_2012}; \citealt{Rasmussen_et_al_2012}). So far, pre-processing and overshooting have been advocated as the most likely explanations for these trends (e.g., \citealt{Balogh_et_al_1999}; \citealt{Wetzel_et_al_2012}) which agrees with our \textsc{gimic} results in the case of cold gas and star formation activity. However, the simulations also predict that a direct interaction with the host and its filaments is sufficiently strong to remove the hot gas of galaxies as far out as 5 r$_{200}$.  Testing this observationally in a direct fashion will be challenging, though, as it is extremely difficult to probe the bulk of the hot baryons around even very local galaxies. 

We finally note that, even though we conclude that it is unlikely for cold gas to be \emph{stripped} from galaxies outside r$_{200}$, this does not mean that it remains wholly unaffected: the removal of hot gas stops the replenishment of cold gas lost through star formation, and will therefore have some indirect impact on the cold gas content and star formation activity of galaxies, although not immediately (see also \citealt{Blanton_et_al_2006} and \citealt{Wilman_et_al_2010}, who show that the fraction of red galaxies depends largely on the local, not the large-scale galaxy density). Likewise, a decrease in star formation will, after a further delay, leave an imprint in the galaxy colour. As well as accurately modelling the removal of hot gas inside r$_{200}$, taking into consideration the loss of gas before even reaching this radius may further improve the accuracy of semi-analytic models of galaxy evolution.

As galaxies move inside r$_{200}$ and fall deep within the potential wells of their hosts, they experience not only increased levels of ram pressure, but also other effects such as tidal stripping and harassment by other galaxies. We have seen in the case of overshot galaxies that these effects have a dramatic impact on their internal properties. In a future paper we explore in detail this complex interplay of environmental influences in the central region of groups and clusters. 

\section*{acknowledgements}
The authors thank the members of the \textsc{gimic} team, particularly Robert Crain, Tom Theuns, and Joop Schaye, for allowing us to use the simulations. They also thank Sean McGee, David Wilman, Robert Crain and Vasily Belokurov for helpful comments on the manuscript.  YMB acknowledges a postgraduate award from STFC. IGM is supported by an STFC Advanced Fellowship. MLB is supported by a NSERC Discovery Grant.  ASF is supported by a Royal Society Dorothy Hodgkin Fellowship. This research has made use of the \textsc{darwin} High Performance Computing Facility at the University of Cambridge. The \textsc{gimic} simulations were carried out using the HPCx facility at the Edinburgh Parallel Computing Centre (EPCC) as part of the EC's DEISA `Extreme Computing Initiative' and with the Cosmology Machine at the Institute of Computational Cosmology of Durham University.

\bibliographystyle{mn2e}
\bibliography{Onset}

\appendix

\end{document}